\documentclass[12pt,psamsfonts,reqno]{amsart}

\theoremstyle{definition}

\theoremstyle{remark}

\newcommand{\vev}[1]{\left\langle #1 \right\rangle}
\newcommand{\ket}[1]{\left |  #1 \right \rangle}
\newcommand{\bra}[1]{\left \langle  #1 \right |}

\newcommand {\CalN} {\mathcal N}

\newcommand {\CalX} {\mathcal X}

\newcommand {\BR}   {\mathbb R}
\newcommand {\BZ}   {\mathbb Z}
\newcommand {\BC}   {\mathbb C}

\newcommand{\bw}{\mathbf{w}}

\newcommand{\bH}{\mathbf{H}}

\newcommand{\bX}{\mathbf{X}}

\newcommand{\bY}{\mathbf{Y}}

\newcommand{\msS}{\mathscr{S}}
\newcommand{\si}{\mathsf{i}}
\newcommand{\sn}{\mathsf{n}}
\newcommand{\sw}{\mathsf{w}}

\newcommand{\sS}{\mathsf{S}}
\newcommand{\sV}{\mathsf{V}}
\newcommand{\sT}{\mathsf{T}}

\newcommand{\sY}{\mathsf{Y}}

\newcommand{\g}{\mathfrak{g}}

\newcommand{\frakM}{\mathfrak{M}}

\newcommand{\fq}{\mathfrak{q}}

\newcommand{\ep}{\epsilon}

\DeclareMathOperator{\Tr} {Tr}

\DeclareMathOperator{\rk} {rk}

\newcommand{\Ind}{\mathbb{I}}

\numberwithin{equation}{section}

\usepackage[margin=1in]{geometry}

\usepackage{amsmath}
\usepackage{amssymb}
\usepackage{amsxtra}
\usepackage{amscd}
\usepackage[mathscr]{euscript}
\usepackage{bbm}

\usepackage[pagebackref=false]{hyperref}
\hypersetup{colorlinks = false, ,extension = notused,linkcolor = blue, anchorcolor = red,citecolor = blue,filecolor = red,pagecolor = red,urlcolor = blue}
\usepackage[numbers,sort&compress]{natbib}
\usepackage{hypernat}

\usepackage{iftex}
\ifxetex
        \usepackage{fontspec}
\else
\fi

\usepackage{float}
\usepackage{tikz-cd}
\usetikzlibrary{decorations.pathreplacing}
\usepackage[mathscr]{euscript}

\begin{document}

\title{Quiver elliptic W-algebras}

\author{Taro Kimura}
\author{Vasily Pestun}

\address{Taro Kimura, Keio University, Japan}
\address{Vasily Pestun, IHES, France} 

\begin{abstract} 
 We define elliptic generalization of W-algebras associated with arbitrary quiver using our construction~\cite{Kimura:2015rgi} with six-dimensional gauge theory.
\end{abstract}

\maketitle 

\tableofcontents

\parskip=4pt

\section{Introduction}

The Seiberg--Witten theory \cite{Seiberg:1994aj,Seiberg:1994rs} provides
an illuminating geometrical description
of the Coulomb branch of the moduli space of vacua
$\mathcal{M}_{\text{vacua}}$ of the
four-dimensional $\CalN=2$ gauge theory: the space
$\mathcal{M}_{\text{vacua}}$ is the base of a certain algebraic integrable
system associated to the 4d $\mathcal{N}=2$ gauge
theory~\cite{Gorsky:1995zq,Martinec:1995by,Donagi:1995cf,Seiberg:1996nz}. The
abelian varieties  appearing in the fibers of the algebraic
integrable system can be realized as Jacobian (Prym) varieties
of certain spectral algebraic curve called Seiberg--Witten curve.  

A typical example of an integrable system associated to $\mathcal{N}=2$
gauge theory is the affine $\sn$-particle Toda model associated to the pure vector
$\mathcal{N}=2$ $U(\sn)$ gauge multiplet. Another example is the classical limit of
$SU(2)$  rational $r$-matrix  spin chain, which is associated to 
 $U(\sn)$ theory with $\sn_{\text{f}} = 2\sn$  fundamental matter
hyper multiplets~\cite{Gorsky:1996hs}.

For  $\mathcal{N}=2$ gauge theory with the gauge group $\times_{i} U(\sn_i)$,
 and with matter hyper multiplets in the fundamental and
bifundamental representations,  encoded by  a
quiver graph $\Gamma$, 
the corresponding integrable system is the integrable system of
monopoles on $\BR^2 \times S^1$ with gauge group $G_{\Gamma}$,  whose
Dynkin diagram is given by the graph $\Gamma$
\cite{Nekrasov:1996cz,Nekrasov:2012xe}. Quantization of this integrable
system is described by means of the Yangian algebra
$\mathbf{Y}(\mathfrak{g}_{\Gamma})$, which is also underlying symmetry
algebra of rational $r$-matrix $\mathfrak{g}_{\Gamma}$ spin chain. 

Similarly, the correspondence holds for 5d $\Gamma$-quiver gauge
theory compactified on $S^1$, or 6d $\Gamma$-quiver gauge theory
compactified on $T^2$: the respective integrable system is the
integrable system of $G_{\Gamma}$-monopoles on $\BR \times T^2$ (integrable $\g_{\Gamma}$-spin chains with trigonometric
$r$-matrix) or $G_{\Gamma}$-monopoles on   $T^3$ (integrable
$\g_{\Gamma}$-spin chains with elliptic $r$-matrix). 

The quantization of the algebraic integrable system with Planck
parameter $\hbar$ corresponds to the equivariant  deformation
$\BR^{4}_{\ep_1, \ep_2}$ of the space-time of the gauge theory
\cite{Moore:1997dj,Nekrasov:2002qd} in the limit
$(\epsilon_1,\epsilon_2) = (\hbar, 0)$, and the supersymmetric vacua of the
gauge theory on $\BR^{4}_{\hbar,0}$ are identified with the spectrum of
quantum integrable system \cite{Nekrasov:2009zz,Nekrasov:2009rc}.

After quantization the classical spectral curve (Seiberg--Witten curve) is
promoted to the Baxter's TQ-relation (for linear chain quivers), or,
more generally, to the equation that states that the $q$-character
\cite{Frenkel:1998} evaluated on a certain element of quantum affine algebra is
a polynomial function of the spectral parameter ~\cite{Nekrasov:2013xda}.
The polynomiality conjecture of \cite{Frenkel:1998} has been proven
in \cite{Frenkel:2013uda}. 

The polynomial equation still holds for generic equivariant parameters
$(\ep_1, \ep_2)$ if the $q$-character is replaced by
a certain algebraic object. This object is called $q_1q_2$-character
in \cite{Nekrasov:2015wsu}, and in \cite{Kimura:2015rgi} this object was
shown to be a generating current of $W_{q_1, q_2}(\g_{\Gamma})$ algebra of
\cite{Frenkel:1997}, a generalization of $q$-Virasoro algebra of \cite{Shiraishi:1995rp} and
\cite{Frenkel:1996}.   (See also \cite{Kim:2016qqs} for realization of
 $qq$-character as a defect partition function).


The gauge theory construction
 naturally involves bifundamental mass parameter $\mu_e$  assigned to each quiver 
edge $e \in \Gamma_1$.  The cohomology class  $[\mu] \in H^{1}(\Gamma, \BC^{\times})$ 
is a non-trivial parameter of the algebra (not considered in
\cite{Frenkel:1997}). 

The construction of the algebra $W_{q_1,q_2}(\mathfrak{sl}_{r+1})$ associated to
quiver gauge theory with $r$ nodes explains the 5d version
\cite{Awata:2009ur} of the AGT relation~\cite{Alday:2009aq} for linear
$sl_{r+1}$-type quivers, with $U(n)^{r}$ gauge group, 
when combined with $r \leftrightarrow n$ duality~\cite{Bao:2011rc} 
 of topological string partition function computed by topological vertex
 \cite{Aganagic:2003db,Maulik_2003,Maulik_2004,Iqbal:2007ii}. 
(See also manifestation of the above duality 
 as a spectral duality of integrable systems~\cite{Mironov:2013xva}.)

In this paper, we generalize our construction~\cite{Kimura:2015rgi} to
the case of elliptic W-algebra using a six-dimensional quiver gauge theory  compactified on a
two-torus $\check T^2$. Equivalently, under the T-duality  for an affine ADE 
quiver $\Gamma$, this gauge theory
is realized on a stack of fractional D3 branes in IIB string theory  on
$\BR^{4} \times T^2 \times \BR^{4}/\tilde \Gamma$. Here $\BR^{4} /
\tilde \Gamma$ is the ADE singularity
 where $\tilde \Gamma \subset SU(2)$ is McKay associated discrete
 subgroup,  and $T^2$ is the dual torus to $\check T^2$; see \cite{Nekrasov:2012xe} for more
details.  The complex integrable system remembers only the complex
structure of the torus $T^2$. We denote by $p = e^{2 \pi \imath \tau}$ the multiplicative
modulus of the underlying elliptic curve, so that as a complex variety
the compactification torus $T^2$ is isomorphic to  $\BC^{\times}/
p^{\BZ}$ or to $\BC/(\BZ + \BZ \tau)$.  The multiplicative spectral parameter $x \in \BC^{\times}/
p^{\BZ}$ is periodic 
\begin{align}
 x \simeq p x
 \label{eq:elliptic_coord}
\end{align}

Our construction is similar to the case of  the 5d gauge
theory compactified on a circle $S^1$~\cite{Kimura:2015rgi}, and thus is
applicable to generic quiver in principle. However, because of the
modular anomalies, we find satisfactory physical interpretation only for
`conformal' quivers, those are the quivers for which the corresponding 4d
$\mathcal{N}=2$ theory is conformal. 

Recently, the elliptic deformation of W-algebra for $A$-type quivers has been
also  discussed in the context of topological string and
supersymmetric gauge theory in~\cite{Iqbal:2015fvd,Nieri:2015dts} and in ~\cite{Mironov:2015thk,Mironov:2016cyq,Mironov:2016yue,Awata:2016riz}.
See also~\cite{Tan:2013xba,Koroteev:2015dja,Koroteev:2016znb,Tan:2016cky} for elliptic generalizations.
Our construction generalizes these results to generic quiver.

\subsection*{Acknowledgements}

The work of TK was supported in part by Keio Gijuku Academic Development Funds, JSPS Grant-in-Aid for Scientific Research (No.~JP17K18090), the MEXT-Supported Program for the Strategic Research Foundation at Private Universities ``Topological Science'' (No.~S1511006), JSPS Grant-in-Aid for Scientific Research on Innovative Areas ``Topological Materials Science'' (No.~JP15H05855), and ``Discrete Geometric Analysis for Materials Design'' (No.~JP17H06462).
VP acknowledges grant RFBR 15-01-04217 and RFBR 16-02-01021. The research of VP on this project has received funding from the European Research Council (ERC) under the European Union's Horizon 2020 research and innovation program (QUASIFT grant agreement 677368).

\section{Elliptic quiver gauge theory}

\subsection{Quiver}

We use the notations of \cite{Kimura:2015rgi}.

Let $\Gamma$ be a quiver with the set of nodes $\Gamma_0$ and the set of
edges $\Gamma_1$. The nodes are typically labelled by $i, j \in
\Gamma_0$.  For an edge $e$ we denote its source node by $s(e)$
and its target node by $t(e)$ or write  $e: i \to j$.

A quiver $\Gamma$ defines $|\Gamma_0| \times |\Gamma_0|$ matrix $(c_{ij})$
\begin{align}
 c_{ij} & = 2 \delta_{ij} - \#(e : i \to j) - \#(e : j \to i) \, ,
 \label{eq:Cartan_mat0}
\end{align}
called the quiver Cartan matrix, which is symmetric.
If there are no loops, all the diagonal elements are equal to 2, and such a matrix defines Kac--Moody algebra $\g(\Gamma)$ with Dynkin diagram $\Gamma$.

\subsection{Elliptic index}
\label{sec:elliptic_index}

The six-dimensional Nekrasov's partition function is defined by the elliptic index 
of the instanton moduli space.
Let $\tau$ be the modulus of the torus on which a six-dimensional theory is compactified, and put $p = e^{2\pi \iota \tau}$.
The elliptic index functor $\Ind_p$ converts additive Chern character class to
the multiplicative elliptic class
\begin{align}
 \Ind_p
 \left[ \sum_{i} x_i \right]
 & =
 \prod_{i} \theta(x_i^{-1};p)
 \label{eq:6d-index}
\end{align}
where the short Jacobi theta function is
\begin{align}
 \theta(x;p)
 & =
 (x;p)_\infty
 (px^{-1};p)_\infty
 =
 \exp
 \left(
  - \sum_{m \neq 0} \frac{x^m}{m(1-p^m)}
 \right)
 \, .
 \label{eq:theta_exp}
\end{align}
We assume $p < 1$ in this paper.
Notice that our conventions are different from
Ref.~\cite{Nekrasov:2013xda}, which uses another version of theta function
\begin{align}
 \theta_1(x;p)
 & =
 \iota p^{\frac{1}{8}} (p;p)_\infty x^{-\frac{1}{2}} \theta(x;p)
 \, .
\end{align}
The index functor behaves under the reflection as follows,
\begin{align}
 \Ind_p \left[ \bX^\vee \right]
 & =
 \begin{cases}
  \displaystyle
  (-1)^{\rk \bX} \left( \det \bX \right)
  \Ind_p \left[ \bX \right]
  & (\theta\text{-version}) \\[.5em]
  \displaystyle
  (-1)^{\rk \bX} \Ind_p \left[ \bX \right]
  & (\theta_1\text{-version})\\
 \end{cases}
 \label{eq:reflection}
\end{align}
because of
\begin{align}
 \theta(x;p) & = (-x) \, \theta(x^{-1};p)
 \, .
 \label{eq:reflection_theta} 
\end{align}
In the 5d limit the two versions correspond to the Dolbeault vs Dirac
conventions. In this paper we will use the Dolbeault convention.\footnote{
The Dirac convention is also often used in the  literature, for example,~\cite{Hollowood:2003cv,Haghighat:2013gba,Benini:2013xpa,Nekrasov:2013xda,Aganagic:2016jmx}, while Refs.~\cite{Saito:2014PRIMS,Gadde:2013ftv,Iqbal:2015fvd,Nieri:2015dts} use the Dolbeault.
}
In the limit $p \to 0$, the 6d index is reduced to the 5d index
\begin{align}
 \lim_{p \to 0} \theta(x^{-1};p) = 1 - x^{-1}
 \, .
\label{eq:5d-index}
\end{align}
For conformal quivers the two versions are equivalent. 

To construct W-algebras from quiver gauge theory, we need to incorporate
the higher time variables~\cite{Marshakov:2006ii}, so that the gauge
theory partition function becomes
a generating function of the observables. The fixed points are labelled
by partitions $(\lambda_{i,\alpha,k})_{\alpha\in[1\ldots \sn_i], \,
  k\in[1\ldots\infty]}$. We introduce a set of variables
\begin{align}
 \CalX_i & =
 \{ x_{i,\alpha,k} \}_{\alpha\in[1\ldots \sn_i], \, k\in[1\ldots\infty]}
 \, , \qquad
 x_{i,\alpha,k} = \nu_{i,\alpha} q_1^{k-1} q_2^{\lambda_{i,\alpha,k}}
 \, , \qquad
 \CalX = \bigsqcup_{i \in \Gamma_0} \CalX_i
\end{align}
where $\sn_i \in \BZ_{\ge 1}$ is the rank of gauge group $U(\sn_i)$ assigned to the node $i \in \Gamma_0$, and $(\nu_{i,\alpha})_{i\in\Gamma_0, \, \alpha\in[1\ldots\sn_i]}$ are the exponentiated Coulomb moduli parameters.
Let $\si: \CalX \to \Gamma_0$ be the node label such that $\si(x) = i$ for $x \in \CalX_i$.
The time variables give rise to the potential term in the partition function
\begin{align}
 \exp
 \left(
 \sum_{m=1}^\infty t_{i,m}^{(\pm)} \bY_i^{[\pm m]}
 \right)
 \label{eq:pot_term}
\end{align}
where $ \bY_i^{[m]}$ are fundamental observables of the quiver gauge
theory~\cite{Nekrasov:2012xe,Nekrasov:2013xda} 
\begin{align}
 \bY_i^{[m]}
 & =
 \frac{1 - q_1^m}{1 - p^m}
 \sum_{x \in \CalX_i} x^m
 \, .
\end{align}
with the notation $\bY_i^{[-m]} = (\bY_i^{[m]})^\vee$.
Notice that we need to introduce twice as many time higher time
variables compared to the 5d case \cite{Kimura:2015rgi}, which is reflected by the extra plus/minus label.
This is the specifics of the elliptic algebras~\cite{Clavelli:1973uk,Saito:2014PRIMS}.

\subsection{Partition function}

The extended partition function $Z_{\mathsf{T}}(t)$ can be computed using the localization formula for the $\sT$-fixed points in the moduli space $\frakM$, characterized by a set of partitions~$\lambda$~\cite{Nekrasov:2002qd}:
\begin{align}
 Z_\sT(t) & =
 \sum_{\CalX \in \frakM^\sT}
 \exp
 \left(
  - \sum_{(x_L,x_R)\in \Lambda^2 \CalX} \sum_{m \neq 0}
  \frac{1 - q_1^m}{m(1 - p^m)(1 - q_2^{-m})}
  \left(c^+_{\si(x_L),\si(x_R)}\right)^{[m]}
  \frac{x_R^m}{x_L^{m}}
 \right)
 \nonumber \\
 & \qquad \qquad \times
 \exp
 \left(
  \sum_{x \in \CalX}
  \left(
   \log \fq_{\si(x)} \log_{q_2} \frac{x}{\mathring{x}}
   + \sum_{m=1}^\infty
   \left(
    \frac{1-q_1^m}{1-p^m} t^{(+)}_{\si(x),m} \, x^m
    + \frac{1-q_1^{-m}}{1-p^{-m}} t^{(-)}_{\si(x),m} \, x^{-m}
   \right)
 \right)
 \right)
\end{align}
where $\fq_i$ is the coupling constant for the node $i \in \Gamma_0$, and $\mathring{x}_{i,\alpha,k} = \nu_{i,\alpha} q_1^{k-1} \in \CalX_0$ denotes the ground configuration corresponding to the empty partition $\lambda=\emptyset$.
The factor $\log_{q_2} \left( x/\mathring{x} \right)$ counts the number of boxes in the partition $\lambda$.
The `positive' part of the Cartan matrix is defined to be
\begin{align}
 c_{ij}^+ & =
 \delta_{ij} - \sum_{e:i \to j} \mu_e^{-1} 
\end{align}
where the multiplicative bifundamental mass parameters $\mu_e \in \BC^\times$ are assigned to  edges $e \in \Gamma_1$, and $\left(c_{ij}^+\right)^{[m]}$ is obtained by replacing $\mu_e$ with $\mu_e^m$, which is the $m$-th Adams operation.
In particular, the vector and bifundamental hypermultiplet contributions are given by
\begin{align}
 Z_i^\text{vec}
 & =
 \prod_{(x,x') \in \CalX_i \times \CalX_i}
 \Gamma\left( q \frac{x}{x'}; q_2, p \right)^{-1}
 \Gamma\left( q_2 \frac{x}{x'}; q_2, p \right)
 \, , \\
 Z_{e:i \to j}^\text{bf}
 & =
 \prod_{(x,x') \in \CalX_i \times \CalX_j}
 \Gamma\left( \mu_e^{-1} q \frac{x}{x'}; q_2, p \right)
 \Gamma\left( \mu_e^{-1} q_2 \frac{x}{x'}; q_2, p \right)^{-1}
 \, ,
\end{align}
where the elliptic gamma function is defined
\begin{align}
 \Gamma(x;p,q) & =
 \exp
 \left(
  \sum_{m \neq 0} \frac{x^m}{m(1-p^m)(1-q^m)}
 \right)
 \, .
\end{align}
Fundamental matter can be realized by a shift of higher time variables.
To add a fundamental factor with mass parameter $\mu \in \BC^\times$ to the node $i \in
\Gamma_0$ one should shift
\begin{align}
 t_{i,m}^{(\pm)}
 \ \longrightarrow \
 t_{i,m}^{(\pm)}
 \pm
 \frac{\mu^{\mp m} q^{\pm m}}{m(1-q_1^{\pm m})(1-q_2^{\pm m})}
 \, .
 \label{eq:t-shift_matter}
\end{align}
As mentioned in Sec.~\ref{sec:V-op}, this shift of $t$-variables is
equivalent to inserting a vertex operator. See also
\cite{Aganagic:2015cta} and \cite{Iqbal:2015fvd}. 

In contrast to 4d and 5d theories, the matter content of 6d theory is restricted due to the gauge anomaly, which is directly related to the modularity of elliptic theory.
The theory is anomaly free if for all nodes $ i \in \Gamma_0$ 
\begin{align}
 c_{ij} \sn_{j} = \sn_{i}^{\text{f}} + \tilde\sn_i^\text{f}
 \label{eq:conf_cond_fund}
\end{align}
where $c_{ij}$ is the Cartan matrix associated with the quiver diagram $\Gamma$ \eqref{eq:Cartan_mat0}, and $\sn_i^\text{f}$ and $\tilde\sn_i^\text{f}$ are the numbers of fundamental and antifundamental multiplets for the node $i$.
The phase space of the algebraic integrable system associated to the Coulomb branch 
of the 6d theory compactified on $\check{T}^{2}$ is the moduli space of
$G_{\Gamma}$-monopoles with singularities on $T^{2} \times S^1$.
The Dirac singularities are associated to the fundamental mass
multiplets, and the total monopole charge is associated to the rank of
the gauge group. 
On a compact space $T^2 \times S^1$, because of the vanishing theorem,
the total monopole charge is necessarily balanced by the charge of Dirac
singularities, and this leads to the balance equation
(\ref{eq:conf_cond_fund}).

We introduce an order $\succ$ on the set of eigenvalues $\CalX$, which is essentially the radial order in CFT.
Then the sum over $(x_L, x_R) \in \Lambda^2 \CalX$ in the partition
function is decomposed into the sum over the pairs $(x_L \succ x_R)$ and
$(x_L \prec x_R)$, and the diagonal terms $(x_L = x_R)$.
Since the diagonal part gives factors independent of the coupling
constant, the Coulomb moduli, and so on, we will omit it. 
Then, using the reflection relation \eqref{eq:reflection}, the partition function is presented
as a sum over the pairs $(x_L \succ x_R)$
\begin{align}
 Z_\sT(t) & =
 \sum_{\CalX \in \frakM^\sT}
 \exp
 \left(
  - \sum_{(x_L \succ x_R)} 
 \left(
  \left(c_{\si(x_L),\si(x_R)}^+\right)^{[0]}
  \beta \, \log \frac{x_R}{x_L} 
  + \sum_{m \neq 0} \frac{1 - q_1^m}{m(1 - p^m)(1 - q_2^{-m})}
  \left(c_{\si(x_L),\si(x_R)}\right)^{[m]}
  \frac{x_R^m}{x_L^{m}}
 \right)
 \right)
 \nonumber \\
 & \qquad \qquad \times
 \exp
 \left(
  \sum_{x \in \CalX}
  \left(
   \log \fq_{\si(x)} \log_{q_2} \frac{x}{\mathring{x}}
   + \sum_{m=1}^\infty
   \left(
    \frac{1-q_1^m}{1-p^m} t^{(+)}_{\si(x),m} \, x^m
    + \frac{1-q_1^{-m}}{1-p^{-m}} t^{(-)}_{\si(x),m} \, x^{-m}
   \right)
 \right)
 \right)
 \label{eq:ext-part-func}
\end{align}
where $\beta = - \epsilon_1 / \epsilon_2$ and the mass deformed Cartan matrix is~\cite{Kimura:2015rgi}
\begin{align}
 c_{ij} = c_{ij}^{+} + c_{ij}^{-}, \quad
 c_{ij}^- = q^{-1}
 (c_{ji}^{+})^\vee, \quad 
 c_{ij} = (1 + q^{-1}) \delta_{ij}
 - \sum_{e:i \to j} \mu_e^{-1}
 - \sum_{e:j \to i} \mu_e q^{-1}
 \, ,
 \label{eq:def-Cartan}
\end{align}
obeying the transpose relation
\begin{align}
 c_{ji} = q^{-1} c_{ij}^\vee
 \, .
 \label{eq:Cartan_inv}
\end{align}
The Cartan matrix \eqref{eq:Cartan_mat0} is reproduced in the limit $q \to 1$, $\mu_e \to 1$.

\section{Operator formalism}

\subsection{$Z$-state}
\label{sec:Z-state}

The gauge theory partition function $ Z_\sT(t)$ \eqref{eq:ext-part-func} can be interpreted  as a state in the Fock space for the infinite-dimensional Heisenberg algebra~$\bH$ generated by $(t_{i,m}^{(\pm)}, \frac{\partial}{\partial t_{i,m}^{(\pm)}})_{i \in \Gamma_0, m \in \BZ_{> 0} }$.
The Fock space is generated from the vacuum state $| 1 \rangle $ by the action of the operators $t_{i,m}^{(\pm)}$.
The vacuum state $ |1 \rangle $ is annihilated by the operators  $\frac{\partial}{\partial t_{i,m}^{(\pm)}}$.

The $Z$-state is presented as an ordered product of vertex operators $
S_{\si(x),x}$ acting on the vacuum
\begin{align}
 \ket{Z_\sT} & =
 \sum_{\CalX \in \frakM^\sT}
 \prod_{x \in \CalX}^\succ
 S_{\si(x),x} \ket{1}
 \, ,
 \label{eq:Z-state}
\end{align}
where 
\begin{align}
 S_{i,x} & = \ :
 \exp
 \left(
 s_{i,0} \log x + \tilde{s}_{i,0}
 + \frac{\kappa_i}{2}
 \left( \log_{q_2}^2 x - \log_{q_2} x \right)
 + \sum_{m \neq 0}
 \left( s_{i,m}^{(+)} x^{-m} + s_{i,m}^{(-)} x^{+m} \right)
 \right) :
\end{align}
The $\kappa_i$ factor, which plays a similar role to the Chern--Simons term, is needed to obtain the agreement between the gauge theory definition and the operator formalism.
As mentioned in Sec.~\ref{sec:elliptic_index}, we need two sets of
time variables for elliptic theory, denoted  by $(t^{(+)}_{i,m})_{i \in \Gamma_0, m \in \BZ_{>0}}$ and $(t^{(-)}_{i,m})_{i \in \Gamma_0, m \in \BZ_{>0}}$.
Then we introduce free field modes 
\begin{align}
 s_{i,-m}^{(\pm)} \stackrel{m>0}{=}
 \frac{1 - q_1^{\pm m}}{1 - p^{\pm m}} t_{i,m}^{(\pm)}
 \, , \quad
 s_{i,0} = t_{i,0} 
 \, , \quad
 s_{i,m}^{(\pm)} \stackrel{m>0}{=}
 \mp \frac{1}{m(1 - q_2^{\mp m})} c_{ji}^{[\pm m]} 
 \frac{\partial}{\partial t_{j,m}^{(\pm)}}
 \, .
\end{align}
The commutation relations are
\begin{align}
 \left[
  s_{i,m}^{(\pm)} \, , \, s_{j,m'}^{(\pm)}
 \right]
 & =
 \mp \frac{1 - q_1^{\pm m}}{m(1 - p^{\pm m})(1 - q_2^{\mp m})}
 c_{ji}^{[\pm m]} \delta_{m+m',0} 
 \, ,
\end{align}
For the zero mode $s_{i,0}$ we set by definition $s_{i,0}^{(\pm)} = s_{i,0}$ and
\begin{align}
 \left[
  \tilde{s}_{i,0} \, , \, s_{j,m}^{(\pm)}
 \right]
 & = - \beta \, \delta_{m,0} \, c_{ji}^{[0]} 
 \, .
\end{align}
In the 5d limit $p \to 0$, the modes $s^{(-)}_{i,m}$ become trivial because
\begin{align}
 - \frac{1}{1 - p^{-m}} = \frac{p^m}{1 - p^m}
 \ \longrightarrow \ 0
 \qquad (p \to 0)
 \, .
\end{align}

The $Z$-state \eqref{eq:Z-state} coincides with the gauge theory definition of the partition function \eqref{eq:ext-part-func} evaluated with the coupling constant
\begin{align}
 \log_{q_2} \fq_i & =
 \beta + t_{i,0} 
 + \sn_j \left( c_{ji}^- \right)^{[\log_{q_2}]} 
 - \log_{q_2} \left( (-1)^{\sn_j} \nu_j \right)  \left( c_{ji}^- \right)^{[0]} 
\end{align}
where
\begin{align}
 \nu_i & =
 \prod_{\alpha=1}^{\sn_i} \nu_{i,\alpha}
 \, , \qquad
 \left( c_{ij}^- \right)^{[\log_{q_2}]} =
 \delta_{ij} \log_{q_2} q^{-1}
 - \sum_{e:j \to i} \log_{q_2} \left( q^{-1} \mu_e \right) 
 \, ,
\end{align}
and the coefficients at all the nodes $i \in \Gamma_0$ satisfy
\begin{align}
 \kappa_i & = \sn_j \left( c_{ji}^- \right)^{[0]} 
 \, .
 \label{eq:conf_cond}
\end{align}

\subsection{Screening charge}
The sum over the set of fixed points $\frakM^\sT$ in the partition
function \eqref{eq:Z-state} can be replaced by the sum over
$\BZ^{\CalX_0}$ (see~\cite{Kimura:2015rgi}): 
\begin{align}
 \ket{Z_\sT} & =
 \sum_{\CalX \in \BZ^{\CalX_0}}
 \prod_{x \in \CalX}^\succ
 S_{\si(x),x} \ket{1}
 \, .
 \label{eq:Z-state2}
\end{align}
Define `the screening charge operator' to be:
\begin{align}
 \sS_{i,\mathring{x}} & =
 \sum_{k \in \BZ} S_{i,q_2^{k} \mathring{x}}
 \, ,
\end{align}
Then the state $ \ket{Z_\sT}$ is obtained by the action of the
ordered product of $\sS_i$  on the vacuum state
\begin{align}
 \ket{Z_\sT} & =
 \prod_{\mathring{x} \in \CalX_0}^\succ
 \sS_{\si(\mathring{x}),\mathring{x}}
 \ket{1}
 \, .
\end{align}
The vacuum $\ket{1}$ of the Heisenberg algebra $\mathbf{H}$ is represented by a constant function of the
 time variables $t_{i,m}^{(\pm)}$, and is annihilated by all the
 `positive' oscillators $s^{(\pm)}_{i,m}$ for $m > 0$.
The dual $\bra{1}$ to the vacuum state is the linear form on the Fock
space represented by the evaluation of the functions of
$t_{i,m}^{(\pm)}$ at  $t_{i,m}^{(\pm)}=0$.

Thus the non-$t$-extended partition function can be presented as the
correlator of the screening charges~\cite{Kimura:2015rgi} (and see also~\cite{Aganagic:2013tta,Aganagic:2014oia,Aganagic:2015cta})
\begin{align}
 Z_\sT(t=0) =
 \vev{1 | Z_\sT}
 & =
 \bra{1}
 \prod_{\mathring{x} \in \CalX_0}^\succ
 \sS_{\si(\mathring{x}),\mathring{x}}
 \ket{1}
 \, .
 \label{eq:Z-conf_block}
\end{align}

\subsection{Trace formula}

Thinking in terms of the $q$-CFT on the torus, the correlator for 6d theory
\eqref{eq:Z-conf_block} can be written in the trace form in terms of the operators of the 5d
theory (up to a normalization factor, which can be absorbed by
redefinition of the gauge coupling constant)
\begin{align}
 Z_\sT(t=0) & = \Tr \left[ p^{L_0}
 \prod_{\mathring{x} \in \CalX_0}^\succ
 \sS_{\si(\mathring{x}),\mathring{x}}^\text{5d}
 \right]
 \, .
 \label{eq:Tr_formula}
\end{align}
Here the trace is taken over the Fock space of $\mathbf{H}$ with respect to the 5d time variables $\{t_{i,m}\}_{i \in \Gamma_0, i\in \BZ_{>0}}$, and the screening charge is also defined with the oscillators used in 5d theory \cite{Kimura:2015rgi}.
The energy operator $L_0$ is 
\begin{align}
 L_0 & =
 \sum_{i \in \Gamma_0} \sum_{m=1}^\infty
 m \, t_{i,m} \frac{\partial}{\partial t_{i,m}}
 \, .
\end{align}
The trace formula \eqref{eq:Tr_formula} can be obtained as follows.

Recall that the screening current correlator which gives the 5d gauge
theory partition function is~\cite{Kimura:2015rgi}
\begin{align}
 \Big< S_{i,x}^\text{5d} S_{j,x'}^\text{5d} \Big>
 & =
 \exp
 \left(
 - \sum_{m=1}^\infty \frac{1}{m} \frac{1-q_1^m}{1-q_2^{-m}}
 c_{ji}^{[m]} \frac{x'^m}{x^m}
 \right)
 \, .
 \label{eq:SS-pair-5d0}
\end{align}
Here we omit the zero modes for brevity. 
There are two options to deform the 5d index computed by
\eqref{eq:5d-index} to the elliptic 6d index computed by~\eqref{eq:6d-index}.

The first option is to modify the oscillator algebra in such a way that the normal
ordering produces the elliptic correlation function, as defined in Sec.~\ref{sec:Z-state},
\begin{align}
 \Big< S_{i,x}^{\mathrm{6d}} S_{j,x'}^{\mathrm{6d}} \Big>
 & =
 \exp
 \left(
 - \sum_{m \neq 0}^\infty \frac{1-q_1^m}{m(1-p^m)(1-q_2^{-m})}
 c_{ji}^{[m]} \frac{x'^m}{x^m}
 \right)
 \, .
 \label{eq:SS-pair-5d}
\end{align}

The second option is to keep the free field oscillator commutation
relations of the 5d theory of the correlator, but change the definition
of the correlation function to the trace as follows 
\begin{align}
 \Big< S_{i,x}^\text{5d} S_{j,x'}^\text{5d} \Big>_\text{torus}
 =
 \Tr \Big[
 p^{L_0} S_{i,x}^\text{5d} S_{j,x'}^\text{5d}
 \Big]
 \, .
 \label{eq:SS-pair-6d}
\end{align}
The proof of the equivalence 
\begin{align}
 \Big< S_{i,x}^{\mathrm{6d}} S_{j,x'}^{\mathrm{6d}} \Big>
 &  =
 \Tr \Big[
 p^{L_0} S_{i,x}^\text{5d} S_{j,x'}^\text{5d}
 \Big]
 \, ,
 \label{eq:SS-paier-eq}
\end{align}
is in the Appendix~\ref{sec:coh_st}. Then the trace formula
\eqref{eq:Tr_formula} follows.

\begin{figure}[t]
 \centering
 \begin{tikzpicture}[thick,scale=.8]

  \draw (0,0) -- (3,0)
  arc [start angle = -90, end angle = 90, x radius = .3, y radius = 1]
  -- (0,2)
  arc [start angle = 90, end angle = -270, x radius = .3, y radius = 1];


  \draw (5,0)
  arc [start angle = -90, end angle = 90, x radius = 1, y radius = 1]
  arc [start angle = 90, end angle = -270, x radius = .3, y radius = 1];

  \draw (-2,0)
  arc [start angle = -90, end angle = 90, x radius = .3, y radius = 1]
  arc [start angle = 90, end angle = 270, x radius = 1, y radius = 1];

  \draw [dotted] (-2,0)
  arc [start angle = -90, end angle = -270, x radius = .3, y radius = 1];

  \node at (-1,1) {$\times$};
  \node at (4,1) {$\times$};

  \node at (-1,-1.5) {$\times$};
  \node at (4,-1.5) {$\times$};  


  \node at (-1.6,-1.5) [left] {$\bra{1}$};
  \node at (4.6,-1.5) [right] {$\ket{1}$};  
  \node at (1.5,-1.5)
  {$\displaystyle \prod_{\mathring{x} \in \CalX_0}^\succ
    \sS_{\si(\mathring{x}),\mathring{x}}^\text{5d} $};

  \node at (7,1) {$=$};

  \begin{scope}[shift={(11,0)}]

   \draw (0,0) -- (3,0)
   arc [start angle = -90, end angle = 90, x radius = 1, y radius = 1]
   -- (0,2)
   arc [start angle = 90, end angle = -270, x radius = .3, y radius = 1];
   
  \draw (-2,0)
  arc [start angle = -90, end angle = 90, x radius = .3, y radius = 1]
  arc [start angle = 90, end angle = 270, x radius = 1, y radius = 1];

  \draw [dotted] (-2,0)
  arc [start angle = -90, end angle = -270, x radius = .3, y radius = 1];

  \node at (-1,1) {$\times$};
  \node at (-1,-1.5) {$\times$};

   \node at (-1.6,-1.5) [left] {$\bra{1}$};
   \node at (1.7,-1.5)
   {$\displaystyle \ket{Z_\sT^\text{5d}}$};
   
  \end{scope}

  \begin{scope}[shift={(0,-6.5)}]

  \draw (-.5,0) -- (1.5,0)
  arc [start angle = -90, end angle = 90, x radius = .3, y radius = 1]
  -- (-.5,2)
  arc [start angle = 90, end angle = -270, x radius = .3, y radius = 1];

  \draw (2.25,0) -- ++(2,0)
  arc [start angle = -90, end angle = 90, x radius = .3, y radius = 1]
  -- (2.25,2)
  arc [start angle = 90, end angle = -270, x radius = .3, y radius = 1];

  \draw (5,0) -- ++(2,0)
  arc [start angle = -90, end angle = 90, x radius = .3, y radius = 1]
  -- (5,2)
  arc [start angle = 90, end angle = -270, x radius = .3, y radius = 1];
   
  \draw (8.5,0)
  arc [start angle = -90, end angle = 90, x radius = 1, y radius = 1]
  arc [start angle = 90, end angle = -270, x radius = .3, y radius = 1];

  \draw (-2,0)
  arc [start angle = -90, end angle = 90, x radius = .3, y radius = 1]
  arc [start angle = 90, end angle = 270, x radius = 1, y radius = 1];

  \draw [dotted] (-2,0)
  arc [start angle = -90, end angle = -270, x radius = .3, y radius = 1];

   \node at (-1.2,1) {$\cdots$};
   \node at (7.8,1) {$\cdots$};

   \node at (-1.2,2.5) {$\cdots$};
   \node at (7.8,2.5) {$\cdots$};   

   \node at (10.5,1) {$=$};
   \node at (10.5,-1.7) {$=$};   

   \node at (-1.2,-1.7) {$\times$};
   \node at (7.8,-1.7) {$\times$};

   \node at (-1.6,-1.7) [left] {$\bra{1}$};
   \node at (8.3,-1.7) [right] {$\ket{1}$};  
   \node at (3.25,-1.7)
   {$\displaystyle \prod_{\mathring{x} \in \CalX_0}^\succ
     \sS_{\si(\mathring{x}),\mathring{x}}^\text{6d} $};

   \draw
   [decorate,decoration={brace,amplitude=10pt,raise=-4pt}]
   (7.8,-.5) -- (-1.3,-.5) node [black,midway,yshift=-0.6pt] {};
   

   \node at (.5,2.5) {$p^{-1}x$};
   \node at (3.25,2.5) {$x$};
   \node at (6,2.5) {$p x$};   
   
  \end{scope}

  \begin{scope}[shift={(13,-6.5)}]
   
   \draw (1,0) -- ++(1.5,0)
   arc [start angle = -90, end angle = 90, x radius = .3, y radius = 1]
   -- (1,2)
   arc [start angle = 90, end angle = -270, x radius = .3, y radius = 1];

   \draw [thick] (1,1)
   circle [x radius = .3, y radius = 1];

   \draw [thick] (2.5,0)
   arc [start angle = -90, end angle = 90, x radius = .3, y radius = 1];

   \draw [thick,dotted] (2.5,0)
   arc [start angle = -90, end angle = -270, x radius = .3, y radius = 1];

   \node at (-1.5,1) {Tr};

   \node at (-.15,1) {$p^{L_0}$ $\times$};

   \draw (-.8,-.5) -- (-1,-.5) -- (-1,2.5) -- (-.8,2.5);
   \draw (3,-.5) -- (3.2,-.5) -- (3.2,2.5) -- (3,2.5);

   \node at (1.3,-1.7) {\eqref{eq:Tr_formula}};

  \end{scope}
  
 \end{tikzpicture}
 \caption{Conformal blocks as the partition function of 5d (top) and 6d theory (bottom). The 6d block has two equivalent expressions.}
 \label{fig:conf_block}
\end{figure}

The physical meaning is as follows. For the 5d gauge theory we use the
cylindrical space-time for the $q$-Toda to compute the partition function, as shown in the top panel of Fig.~\ref{fig:conf_block}.
For the 6d gauge theory we use the toric space-time for the $q$-Toda obtained by the identification~\eqref{eq:elliptic_coord}, illustrated in the LHS of Fig.~\ref{fig:conf_block} (bottom).
This corresponds to \eqref{eq:Z-conf_block}, and is actually equivalent to taking the trace with the operator $p^{L_0}$ inserted.
This trace version~\eqref{eq:Tr_formula}  also agrees 
with the spectral duality for elliptic theory~\cite{Mironov:2016cyq,
  Iqbal:2015fvd, Nieri:2015dts} because the dual theory is $\CalN=2^*$
theory (or cyclic quiver theory), whose partition function is given by
the torus conformal block via the $q$-version of the AGT relation~\cite{Alday:2009aq}.

\subsubsection{Connection to elliptic quantum group}

 It has been known that the $q$-deformation of W-algebra has a close connection with the elliptic quantum algebra $U_{q,p}(\widehat{\mathfrak{g}})$:
 The screening current of $W_{q_1,q_2}(\mathfrak{g})$ obeys essentially the same relation to the elliptic currents $e_i(z)$ and $f_i(z)$ of $U_{q,p}(\widehat{\mathfrak{g}})$~\cite{Feigin:1995sf}.
See \cite{Farghly:2015ART} for the relations for generic $\mathfrak{g}$.
We see from \eqref{eq:SS-pair-5d0} that the 5d screening currents yields
\begin{align}
 S_{i,x}^\text{5d} S_{j,x'}^\text{5d}
 & = S_{j,x'}^\text{5d} S_{i,x}^\text{5d} \times \exp
 \left(
 - \sum_{m \neq 0} \frac{1}{m} \frac{1 - q_1^m}{1 - q_2^{-m}}
 c_{ji}^{[m]} \left( \frac{x'}{x} \right)^m
 \right)
\end{align}
where we omitted the zero mode factors for simplicity.
One can rewrite the OPE factor using the theta function \eqref{eq:theta_exp}.
Swapping $q_1 \leftrightarrow q_2$ corresponds to swapping the currents $e_i(z) \leftrightarrow f_i(z)$.

From \eqref{eq:SS-pair-5d}, on the other hand, we obtain exactly the same relation for the 6d screening currents
\begin{align}
 S_{i,x}^\text{6d} S_{j,x'}^\text{6d}
 & = S_{j,x'}^\text{6d} S_{i,x}^\text{6d} \times \exp
 \left(
 - \sum_{m \neq 0} \frac{1}{m} \frac{1 - q_1^m}{1 - q_2^{-m}}
 c_{ji}^{[m]} \left( \frac{x'}{x} \right)^m
\right) 
\end{align}
This coincidence implies that both the $q$-deformation $W_{q_1,q_2}(\mathfrak{g})$ and the elliptic deformation $W_{q_1,q_2,p}(\mathfrak{g})$ belong to the same realization of the elliptic quantum algebra $U_{q,p}(\widehat{\mathfrak{g}})$.

\subsection{$\sY$-operator}

To construct W-algebras we introduce  $\sY$-operators, corresponding to the doubled potential term \eqref{eq:pot_term}
\begin{align}
 \sY_{i,x} & = \
 q_1^{\tilde{\rho}_i}
 :
 \exp
 \left(
 y_{i,0} - \left( \tilde{c}_{ji} \right)^{[0]} \kappa_j \log x
 + \sum_{m \neq 0}
 \left( y_{i,m}^{(+)} x^{-m} + y_{i,m}^{(-)} x^{+m} \right)
 \right)
 :
 \, ,
\end{align}
where $\tilde{\rho}_i$ is the Weyl vector defined by $\tilde{\rho}_i =
\sum_{j \in \Gamma_0} \tilde{c}_{ji}^{[0]}$, and $\tilde{c}_{ij}$ is the
inverse of mass-deformed Cartan matrix $c_{ij}$.
The affine case with $\det (c_{ij}^{[0]}) = 0$ will be discussed in Sec.~\ref{sec:A0hat}.
In the following, we set $\kappa_i = 0$ for $\forall i \in \Gamma_0$ for simplicity.

The oscillators $y_{i,m}^{(\pm)}$ are defined in terms of
$t_{i,m}^{(\pm)}$ and $\frac{\partial}{\partial t_{i,m}^{(\pm)}}$
\begin{align}
 y_{i,-m}^{(\pm)} & \stackrel{m > 0}{=}
 \frac{(1-q_1^{\pm m})(1-q_2^{\pm m})}{1-p^{\pm m}}
 \left( \tilde{c}^{[\mp m]} \right)_{ji} t_{j,m}^{(\pm)} 
 \, , \\
 y_{i,m}^{(\pm)} & \stackrel{m > 0}{=}
 \mp \frac{1}{m} \frac{\partial}{\partial t_{i,m}^{(\pm)}}
 \, , \\
 y_{i,0} & =
 - t_{j,0} \tilde{c}_{ji}^{[0]} \log q_2 
 \, .
\end{align}
They satisfy the commutation relation:
\begin{align}
 \left[
  y_{i,m}^{(\pm)} \, , \, y_{j,m'}^{(\pm)}
 \right]
 & =
 \mp \frac{1}{m}
 \frac{(1-q_1^{\pm m})(1-q_2^{\pm m})}{1-p^{\pm m}}
 (\tilde{c}^{[\mp m]})_{ij} \, \delta_{m+m',0} 
 \, .
 \label{eq:yy_osci_comm}
\end{align}
In terms of the free field $s_{i,m}^{(\pm)}$, we have 
\begin{align}
 y_{i,m}^{(\pm)} & \stackrel{m \neq 0}{=}
 (1 - q_2^{\mp m})
 \tilde{c}_{ji}^{[\pm m]} s_{j,m}^{(\pm)} 
 \, , \qquad
 y_{i,0} =
 \left( \log q_2^{-1} \right)
 \tilde{c}_{ji}^{[0]} s_{j,0} 
 \, ,
\end{align}
hence the $[y,s]$ commutation relations are 
\begin{align}
 \left[
  y_{i,m}^{(\pm)} \, , \, s_{j,m'}^{(\pm)}
 \right]
 & =
 \mp \frac{1}{m} \frac{1 - q_1^{\pm m}}{1 - p^{\pm m}}
 \delta_{m + m',0} \delta_{ij}
 \, , \qquad
 \left[
  \tilde{s}_{i,0} \, , \, y_{j,0}
 \right]
 = - \delta_{ij} \log q_1
 \, .
\end{align}
This  leads to the normal ordered product (with the ordering $|x| > |x'|$)
\begin{align}
 \sY_{i,x} S_{i,x'}
 & =
 \frac{\theta(x'/x;p)}{\theta(q_1x'/x;p)}
 : \sY_{i,x} S_{i,x'} : \, ,
 \quad
 \sY_{i,x} S_{j,x'} = \ : \sY_{i,x} S_{j,x'} :
 \quad \text{for}  \quad i \neq j
 \, .
 \label{eq:prod_YS}
\end{align}
The expectation value of the $\sY$-function has infinitely many poles at
$x = x' q_1 p^n$ for $\forall n \in \BZ$ for each configuration
$\mathcal{X} \in \mathfrak{M}^{\sT}$ that labels the insertion of the
screening currents:
\begin{align}
 \bra{1} \sY_{i,x} \prod_{x' \in \CalX}^\succ S_{\si(i),x'} \ket{1}
 & =
 q_1^{\tilde{\rho}_i}
 \left(
 \prod_{x' \in \CalX_i}
 \frac{\theta(x'/x;p)}{\theta(q_1 x'/x;p)}
 \right)
 \bra{1} \prod_{x' \in \CalX}^\succ S_{\si(i),x'} \ket{1}
 \, .
\end{align}
On the other hand, for $|x| < |x'|$, we have
\begin{align}
 S_{i,x'} \sY_{i,x} 
 & =
 q_1^{-1} \frac{\theta(x/x';p)}{\theta(q_1^{-1}x/x';p)}
 : \sY_{i,x} S_{i,x'} : 
 \, .
\end{align}
Therefore the commutator gives
\begin{align}
 \left[
  \sY_{i,x} \, , \, S_{i,x'}
 \right]
 & =
 \left(
 \frac{\theta(x'/x;p)}{\theta(q_1x'/x;p)}
 - q_1^{-1} \frac{\theta(x/x';p)}{\theta(q_1^{-1}x/x';p)}
 \right)
 : \sY_{i,x} S_{i,x'} :
 \nonumber \\
 & =
 \frac{\theta(q_1^{-1};p)}
      {(p;p)_\infty^2} \,
 \delta(q_1 x'/x)
 : \sY_{i,x} S_{i,x'} :
 \, .
\end{align}
The last expression is due to the identity~\cite{Saito:2014PRIMS}
\begin{align}
 \frac{\theta(a z;p)}{\theta(z;p)}
 & =
 \frac{\theta(a;p)}{(p;p)_\infty^2}
 \sum_{n \in \BZ} \frac{z^n}{1 - a p^n}
 \, ,
\end{align}
which is obtained by using Ramanujan's summation formula with the delta function defined
\begin{align}
 \delta(x) = \sum_{n \in \BZ} x^n \, .
\end{align}
This means that, in the limit $q_1 \to 1$, the $\sY$-operator commutes with the screening current, and it reproduces a commutative algebra~\cite{Nekrasov:2013xda}.


\subsection{$\sV$-operator}\label{sec:V-op}

We can incorporate the (anti)fundamental matter contribution in the operator formalism by considering another vertex operator,
\begin{align}
 \sV_{i,x}
 & =
 \ : \exp
 \left(
  \sum_{m \neq 0}
  \left( v_{i,m}^{(+)} x^{-m} + v_{i,m}^{(-)} x^{+m} \right)
 \right):
 \, .
\end{align}
To reproduce the $t$-variable shift \eqref{eq:t-shift_matter}, the oscillators are taken to be
\begin{align}
 v_{i,-m}^{(\pm)} & \stackrel{m > 0}{=}
 - \frac{1}{1-p^{\pm m}}
 \tilde{c}_{ji}^{[\pm m]} t_{j,m}^{(\pm)}
 \, , \qquad
 v_{i,m}^{(\pm)} \stackrel{m > 0}{=}
 \pm \frac{1}{m}
 \frac{1}{\left(1-q_1^{\pm m}\right)\left(1-q_2^{\pm m}\right)}
 \frac{\partial}{\partial t_{i,m}^{(\pm)}}
 \, ,
\end{align}
and the commutation relation
\begin{align}
 \left[
  v_{i,m}^{(\pm)} \, , \, s_{j,m'}^{(\pm)}
 \right]
 & =
 \pm \frac{1}{m(1-p^{\pm m})(1-q_2^{\pm m})} \,
 \delta_{m+m',0} \delta_{ij}
 \, .
\end{align}
We remark
\begin{align}
 v_{i,m}^{(\pm)} & =
 - \frac{1}{\left(1-q_1^{\pm m}\right)\left(1-q_2^{\pm m}\right)}
 y_{i,m}^{(\pm)}
 \, .
\end{align}
The product of $\sV$ and $\sS$ operators behaves
\begin{align}
 \sV_{i,x} \sS_{i,x'}
 & =
 \Gamma\left( \frac{x'}{x}; p, q_2 \right)
 : \sV_{i,x} \sS_{i,x'} :
 \, , \\
 \sS_{i,x'} \sV_{i,x} 
 & =
 \Gamma\left( \frac{x}{x'}; p, q_2 \right)^{-1}
 : \sV_{i,x} \sS_{i,x'} :
 \, ,
\end{align}
which corresponds to the fundamental and antifundamental matter factors, respectively, while the OPE of $\sV$ and $\sV$ does not yield dynamical contribution.
The $t$-extended partition function with the (anti)fundamental matter factors is given by
\begin{align}
 \ket{Z_\sT} & =
 \left(
 \prod_{x \in \CalX_\text{f}}
 \sV_{\si(x),x}
 \right)
 \left(
 \prod_{\mathring{x} \in \CalX_0}^\succ
 \sS_{\si(\mathring{x}),\mathring{x}}
 \right)
 \left(
 \prod_{x \in \tilde\CalX_\text{f}}
 \sV_{\si(x),x}
 \right)
 \ket{1}
 \label{eq:Z-state_matter}
\end{align}
where 
$\CalX_\text{f} = \{\mu_{i,f}\}_{i\in\Gamma_0,f\in[1,\ldots,\sn_i^\text{f}]}$
and
$\tilde\CalX_\text{f} = \{\tilde\mu_{i,f}\}_{i\in\Gamma_0,f\in[1,\ldots,\tilde\sn_i^\text{f}]}$
are sets of (anti)fundamental mass parameters.
This mass parameter characterizes the pole on the elliptic curve at $x = \mu_{i,f}$, which is added by the vertex operator $\sV_{\si(x),x}$ with $x \in \CalX_\text{f}$ and $\tilde\CalX_\text{f}$.

We again remark that, for the modular invariance of the non-extended partition function $\vev{1 | Z_\sT}$, which is a conformal block of $W(\Gamma)$-algebra, we have to take into account the conformal condition \eqref{eq:conf_cond_fund}, although the $Z$-state \eqref{eq:Z-state_matter} is not necessarily modular invariant by itself.

\section{Elliptic W-algebra}

To construct quiver elliptic W-algebras we build holomorphic
$qq$-character currents~\cite{Nekrasov:2015wsu,Kimura:2015rgi} (see also
\cite{Bourgine:2015szm,Bourgine:2016vsq}). 

\subsection{$A_1$ quiver}

For the simplest quiver $\Gamma = A_1$, the $qq$-character is
\begin{align}
 T_{1,x} & = \sY_{1,x} + \sY_{1,q^{-1}x}^{-1}
 \, .
\end{align}
Let us show that  $T_{1,x}$ commutes with the screening charge $\sS_{1,x'}$, which assures the regularity of the $qq$-character.

Here are the possible terms appearing in the commutation relation between the $qq$-character and the screening current,
\begin{align}
 \sY_{1,x} S_{1,x'}
 & =
 \frac{\theta(x'/x;p)}{\theta(q_1 x'/x;p)}
 : \sY_{1,x} S_{1,x'} :
 \, , \label{eq:YS_A1_1} \\
 S_{1,x'} \sY_{1,x}
 & =
 q_1^{-1} \frac{\theta(x/x';p)}{\theta(q_1^{-1} x/x';p)}
 : S_{1,x'} \sY_{1,x} :
 \, , \label{eq:YS_A1_2} \\
 \sY_{1,q^{-1}x}^{-1} S_{1,x''}
 & =
 \frac{\theta(qq_1 x''/x;p)}{\theta(qx''/x;p)}
 : \sY_{1,q^{-1}x}^{-1} S_{1,x''} :
 \, , \label{eq:YS_A1_3} \\
 S_{1,x''} \sY_{1,q^{-1}x}^{-1}
 & =
 q_1
 \frac{\theta(q^{-1}q_1^{-1} x/x'';p)}{\theta(q^{-1}x/x'';p)}
 : S_{1,x''} \sY_{1,q^{-1}x}^{-1} :
 \, . \label{eq:YS_A1_4}
\end{align}
As shown in \eqref{eq:prod_YS}, the first two terms possibly have infinitely many poles at $x = x' q_1 p^n$ for $n \in \BZ$, while the last two terms involve poles at $x = x'' q p^n$.
These poles are actually cancelled with each other because the screening charge is defined as a sum over the the screening current under the $q_2$-shift: there is a term $S_{1,x''}$ with $x' = q_2 x''$ for every $S_{1,x'}$.
Using the relation
\begin{align}
 q_1^{-1}
 :
 \sY_{1,x} \sY_{1,q^{-1} x}
 :
 \ = \
 :
 S_{1,q^{-1} x} S_{1,q_1^{-1} x}^{-1}
 :
 \, ,
\end{align}
we can show that the residue of the first term \eqref{eq:YS_A1_1} at $x = x' q_1 p^n$ coincides with that of the fourth term \eqref{eq:YS_A1_4} at $x = x'' q p^n$ for $\forall n \in \BZ$ with $x' = x'' q_2$,
\begin{align}
 \underset{x \to x' q_1 p^n}{\operatorname{Res}}
 \Big[ \sY_{1,x} S_{1,x'} \Big]
 & =
 \underset{x \to x' q_1 p^n}{\operatorname{Res}}
 \left[ S_{1,x' q_2^{-1}} \sY_{1,xq^{-1}}^{-1} \right]
 \, .
\end{align}
Similarly we have a coincidence of \eqref{eq:YS_A1_2} and \eqref{eq:YS_A1_3},
\begin{align}
 \underset{x \to x' q_1 p^n}{\operatorname{Res}}
 \Big[ S_{1,x'} \sY_{1,x} \Big]
 & =
 \underset{x \to x' q_1 p^n}{\operatorname{Res}}
 \left[ \sY_{1,xq^{-1}}^{-1} S_{1,x' q_1^{-1}} \right] 
 \, .
\end{align}
This shows that the regularity of the $qq$-character for elliptic $A_1$ theory
\begin{align}
 \partial_{\bar{x}} T_{1,x} \ket{Z_\sT} & = 0
 \, ,
\end{align}
which is equivalent to the commutativity of the holomorphic current with the screening charge,
\begin{align}
 \Big[ T_{1,x} \, , \, \sS_{1,x'} \Big] & = 0
 \, .
\end{align}
The commutant of the screening charge is a well-defined conserved current, which provides time-independent modes
\begin{align}
 T_{1,x} & = \sum_{m \in \BZ} T_{1,m} \, x^{-m}
 \, .
\end{align}
The modes of this conserved current define the elliptic algebra
$W_{q_1,q_2,p}(A_1)$ as the subalgebra of the Heisenberg algebra
$\bH$ (the $A_1$ case was discussed in~\cite{Iqbal:2015fvd} and~\cite{Nieri:2015dts}).

\subsection{Higher weight current and collision}

We can compute the higher weight current $T^{[\bw]}_{i,x}$ using the free field representation of the vertex operator.
For $A_1$ quiver, the degree two current plays an important role in the definition of the elliptic algebra $W_{q_1,q_2,p}(A_1)$.
The product is given by
\begin{align}
 T_{1,w_1 x} T_{1,w_2 x}
 & =
 \left(
  \sY_{1,w_1 x} + \sY_{1,w_1 x q^{-1}}^{-1}
 \right)
 \left(
  \sY_{1,w_2 x} + \sY_{1,w_2 x q^{-1}}^{-1}
 \right)
 \nonumber \\
 & =
 f(w_2/w_1)^{-1}
 \Big(
 : \sY_{1,w_1 x} \sY_{1,w_2 x} :
 \nonumber \\
 & \qquad
 + \msS(w_1/w_2) : \sY_{1,w_1x} \sY_{1,w_2xq^{-1}}^{-1} :
 + \, \msS(w_2/w_1) : \sY_{1,w_2x} \sY_{1,w_1xq^{-1}}^{-1} :
 \nonumber \\
 & \hspace{20em}
 + : \sY_{1,w_1 x q^{-1}}^{-1} \sY_{1,w_2 x q^{-1}}^{-1} :
 \Big)
\end{align}
where the scalar factor
\begin{align}
 f(w) & =
 \exp
 \left(
  \sum_{m \neq 0}
  \frac{(1 - q_1^m)(1 - q_2^m)}{m (1 - p^m)(1 + q^m)}
  w^m
 \right)
\end{align}
which appears in the algebraic relation of the elliptic Virasoro algebra $W_{q_1,q_2,p}(A_1)$~\cite{Nieri:2015dts}, and the permutation factor
\begin{align}
 \msS(w) & =
 \frac{\theta(q_1 w;p) \theta(q_2 w;p)}{\theta(w;p) \theta(q w;p)}
 \, .
 \label{eq:permutation-fac}
\end{align}
Notice the relation
\begin{align}
 f(w) f(qw) & = \msS(w)
 \, .
\end{align}
We obtain the commutation relation for the elliptic Virasoro generators,
defined as the modes of the holomorphic current, from the product expression
\begin{align}
 &
 f(w_2/w_1) T_{1,w_1 x} T_{1,w_2 x}
 - f(w_1/w_2) T_{1,w_2 x} T_{1,w_1 x}
 \nonumber \\
 & \hspace{10em}
 =
 \frac{\theta(q_1;p) \theta(q_2;p)}{(p;p)_\infty^2 \theta(q;p)}
 \left(
  \delta \left( q \frac{w_1}{w_2} \right)
  - \delta \left( q \frac{w_2}{w_1} \right)
 \right)
 \, .
\end{align}

The degree $n$ current with the weight $(w_1, \ldots, w_n)$ is  computed in a similar way
\begin{align}
 T_{1,x}^{[\bw]} & = \, 
 :\sY_{1,w_1 x} \sY_{1,w_2 x} \cdots \sY_{1,w_n}: + \cdots
 \nonumber \\
 & =
 \sum_{I \cup J = \{1\ldots n\}}
 \prod_{i \in I, j \in J} 
 \msS \left( \frac{w_i}{w_j} \right)
 :\prod_{i \in I} \sY_{1,w_i x}
 \prod_{j \in J} \sY_{1,w_j x q^{-1}}^{-1}:
 \, .
\end{align}
Algebraically, this expression is exactly the same as in the
$W_{q_1,q_2}(A_1)$-algebra~\cite{Nekrasov:2015wsu,Kimura:2015rgi},
except that the factor $ \msS(w)$ is replaced by the elliptic index
version. 

In addition, we can consider the collision limit of the currents which
produces derivative terms. 
For example, the product of the currents $\sY_{1,x} \sY_{1,w x}$ in the
limit $w \to 1$ gives 
\begin{align} 
 &
 \msS \left( w \right)
 : \sY_{1,w x} \sY_{1,x q^{-1}}^{-1} :  
 + \, \msS \left( w^{-1} \right)
 : \sY_{1,x} \sY_{1,w x q^{-1}}^{-1} :
 \nonumber \\
 & \stackrel{w \to 1}{\longrightarrow} \
 : \sY_{1,x} \sY_{1,x q^{-1}}^{-1} 
 \left(
 \mathfrak{c}(q_1,q_2,p) 
 - \frac{\theta(q_1;p) \theta(q_2;p)}{(p;p)_\infty^2\theta(q;p)}
 x \partial_x \log \left( \sY_{1,x} \sY_{1,xq^{-1}} \right)
 \right) :
 \, .
\end{align}
The coefficient is given by
\begin{align}
 \mathfrak{c}(q_1,q_2,p) & =
 \lim_{w \to 1}
 \left( \msS(w) + \msS(w^{-1}) \right)
 \, ,
\end{align}
which is finite, although the factor $\msS(w)$ itself has a pole at $w=1$.

\subsection{Generic quiver}

Now we have an algorithm to compute the $qq$-character for generic quiver.
The fundamental $qq$-character corresponding to the node $i \in
\Gamma_0$  starts with $\sY_{i,x}$, and  the following terms are
generated by the iWeyl reflection~\cite{Nekrasov:2012xe,Nekrasov:2013xda,Nekrasov:2015wsu,Kimura:2015rgi}:
\begin{align}
\label{eq:Tixlocal}
 T_{i,x} & = \sY_{i,x} \, + : \sY_{i,xq^{-1}}^{-1}
 \prod_{e:i \to j} \sY_{j,\mu_e^{-1} x}
 \prod_{e:j \to i} \sY_{j,\mu_e q^{-1} x}
 : + \cdots
 \, .
\end{align}
%
%
If there is a product of the same $\sY$-operators in a term, we apply the permutation factor \eqref{eq:permutation-fac}.
We can prove the regularity of this current, and the commutativity with
the screening charge in exactly the same way as in~\cite{Kimura:2015rgi},
\begin{align}
 \Big[ T_{i,x} \, , \, \sS_{i,x'} \Big]  & = 0
 \, .
\end{align}
Similarly, the holomorphic current $T_{i,x}$ generates well-defined modes, which define the elliptic algebra $W_{q_1,q_2,p}(\Gamma)$.

\subsection{$A_r$ quiver}
We consider an example of a linear quiver $A_r$, which leads to the elliptic algebra $W_{q_1,q_2,p}(A_r)$.
The iWeyl reflection computes the $qq$-character for each node $i \in \Gamma_0$,
\begin{align}
 T_{i,\mu_{1 \Rightarrow i}^{-1} x} & =
 \sum_{1 \le j_1 < \cdots < j_i \le r+1}
 :
 \prod_{k=1}^i \Lambda_{j_k,q^{-i+k}x}
 :
\end{align}
where
\begin{align}
 \Lambda_{i,x} & =
 \sY_{i,\mu_{1 \Rightarrow i}^{-1} \, x}
 \sY^{-1}_{i-1, \mu_{1 \Rightarrow i-1}^{-1} q^{-1} x }
\end{align}
with $\sY_{0,x} = \sY_{r+1,x} = 1$ and the mass product defined
\begin{align}
 \mu_{1 \Rightarrow i}
 & :=
 \mu_{1 \to 2} \mu_{2 \to 3} \cdots \mu_{i-1 \to i}
 = \prod_{j=1}^{i-1} \mu_{j \to j+1}
 \, .
\end{align}
Remark that
\begin{align}
 :
 \prod_{i=1}^{r+1}
 \Lambda_{i,q^{i-1}x}
 :
 & = 1
 \, .
\end{align}
This is the mass deformed quantum Miura transformation of $A_r$ theory.
For example, the first character, corresponding to the fundamental representation, is given by
\begin{align}
 T_{1,x} & =
 \sum_{i=1}^{r+1} \Lambda_{i,x}
 \nonumber \\
 & =
 \sY_{1,x}
 + 
 :
 \sY_{2,\mu_{1 \Rightarrow 2}^{-1} x} \sY_{1,q^{-1} x}^{-1}
 :
 + \cdots
 + 
 :
 \sY_{r,\mu_{1 \Rightarrow r}^{-1} q^{-1} x}^{-1}
 :  
 \, .
\end{align}

\subsection{$\widehat{A}_0$ quiver}\label{sec:A0hat}

So far we have assumed that there is no self-connecting  edge (a loop) in a quiver.
Let us now study an example having a single node with a self-connecting
edge, called $\widehat{A}_0$ quiver (corresponding to 4d $\CalN=2^*$ gauge theory), whose Cartan matrix is $(0)$ and its mass-deformation is given by~\cite{Kimura:2015rgi}
\begin{align}
 c = 1 + q^{-1} - \mu^{-1} - \mu q^{-1}
 \, .
\end{align}
where $\mu \in \BC^\times$ is the multiplicative adjoint mass. 
Then the commutation relation for the free field $y_{i,m}^{(\pm)}$ \eqref{eq:yy_osci_comm} becomes
\begin{align}
 \left[
  y_{1,m}^{(\pm)} \, , \, y_{1,m'}^{(\pm)}
 \right]
 & =
 \mp \frac{1}{m}
 \frac{(1-q_1^{\pm m})(1-q_2^{\pm m})}
      {(1-p^{\pm m})(1-\mu^{\pm m})(1-\mu^{\mp m}q^{\pm m})}
 \delta_{m+m',0}
 \, .
 \label{eq:free_field_A0}
\end{align}
The generating current is obtained from the iWeyl reflection
\begin{align}
 \sY_{1,x}
 + \fq \, \msS(\mu^{-1}) \,
 : \sY_{1,xq^{-1}}^{-1} \sY_{1,\mu^{-1} x}\sY_{1,\mu q^{-1} x} :
 \, ,
\end{align}
which provides a closed formula
\begin{align}
 T_{1,x} & =
 \sum_{\lambda} \fq^{|\lambda|}
 Z_\lambda^{\hat{A}_0}
 (\tilde{q}_1,\tilde{q}_2,\tilde{\nu},\tilde{\mu})
 :
 \prod_{s \in \partial_+ \lambda}
 \sY_{1,qx/\tilde{x}(s)} 
 \prod_{s \in \partial_- \lambda}
 \sY_{1,x/\tilde{x}(s)}^{-1}
 :
 \, ,
 \label{eq:qq-char_A0}
\end{align}
where $\partial_+ \lambda$ and $\partial_- \lambda$ are the outer and inner boundary of the partition $\lambda$, and we define
\begin{align}
 \tilde{x}(s)
 & =
 (\mu^{-1} q)^{s_1-1} \mu^{s_2-1} q
 \, .
 \label{eq:dual-eigenvalue}
\end{align}
The factor $Z_\lambda^{\hat{A}_0}
(\tilde{q}_1,\tilde{q}_2,\tilde{\nu},\tilde{\mu})$ is the elliptic
Nekrasov function for $\hat{A}_0$ quiver ($\CalN=2^*$ theory) with
$U(1)$ gauge node, evaluated at the partition $\lambda$ with the
``dual'' variables~\cite{Nekrasov:2013xda,Kimura:2015rgi}
\begin{align}
 \tilde{q}_1 = \mu^{-1} q
 \, , \qquad
 \tilde{q}_2 = \mu
 \, , \qquad
 \tilde{\mu} = q_2
 \, , \qquad
 \tilde{\nu} = q
 \, ,
\end{align}
so that \eqref{eq:dual-eigenvalue} leads to the dual of $x$-variable $\tilde{x}(k,\lambda_k+1) = \tilde{x}_k \in \tilde{\CalX}$ where $(k,\lambda+1) \in \partial_+ \lambda$.

The current \eqref{eq:qq-char_A0} is an infinite sum over partitions. Notice
that for degenerate Cartan matrix it is not possible to absorb the
zero mode coupling constant $\mathfrak{q}$ to the definition of the $\sY$, and
$\mathfrak{q}$ appears explicitly in the expression for $T_{1,x}$. For
$|\mathfrak{q}| < 1$ the infinite sum over the partitions converges by
the same argument as the convergence of the generating function for the
number of partitions $\sum_{k=0}^{\infty} n_k \mathfrak{q}^{k} =
\prod_{p=1}^{\infty} (1-\mathfrak{q}^p)^{-1}$. 
We define the affine elliptic algebra $W_{q_1,q_2,p}(\hat{A}_0)$ as
generated by the modes $\tilde T_{1,n}$ of  current $T_{1,x} = \sum_{n=-\infty}^{\infty} \tilde
T_{1,n} x^{-n}$. 
For a  current $T_{1,x}^{[\sw]}$ of higher weight $\sw_1$ the sum is labeled by
$\sw$-colored partitions~\cite{Nekrasov:2015wsu,Kimura:2015rgi}.

\appendix
\section{Proof of trace formula \eqref{eq:SS-paier-eq}}
\label{sec:coh_st}

In this Appendix we prove the equivalence \eqref{eq:SS-paier-eq} using the coherent state basis.

\subsection{Coherent state basis}

The argument in this part is essentially parallel to the textbook~\cite{GSW:ST1987}.

For oscillator algebra generated by $(t, \partial_{t})$ with
$[\partial_{t}, t]= 1$ we consider the coherent state basis in the Fock space
\begin{align}
 \ket{n} = \frac{t^n}{\sqrt{n!}} \ket{0}
 \, , \qquad
 \bra{n} = \bra{0} \frac{\partial^n}{\sqrt{n!}}
 \, , \qquad 
 |z) = e^{zt} \ket{0}
 \, , \qquad
 (z| = \bra{0} e^{z^* \partial}
 \label{eq:basis}
\end{align}
The normalization is 
\begin{align}
 \vev{n|m} = \delta_{n,m}
 \, , \qquad
 (z|w) = e^{z^* w}
 \, .
\end{align}
The states in \eqref{eq:basis} are eigenstates of the filling number operator $t \partial \ket{n} = 
n \ket{n}$ and the annihilation/creation operators $\partial |z) = z |z)$, $(z| t = (z| z^*$.
Notice that the operator $a^{t \partial_t}$ acts on the states $|z)$ and $(z|$ as,
\begin{align}
 a^{t \partial_t} |z) & = |az)
 \, , \qquad
 (z| a^{t \partial_t} = (a^* z|
 \, .
\end{align}

The identity operator can be expressed in terms of the coherent state basis:
\begin{align}
 \mathbbm{1} & =
 \frac{1}{\pi} \int d^2 z \,
 |z) e^{-|z|^2} (z|
\end{align}
where
\begin{align}
 \left< n | \mathbbm{1} | m \right> = \delta_{n,m}
 \, ,
\end{align}
so that the trace of an operator is 
\begin{align}
 \Tr \mathcal{O}
 & =
 \frac{1}{\pi} \int d^2 z \, e^{-|z|^2} (z| \mathcal{O} |z)
 \, .
\end{align}
Then we find~\cite{Yamada:2006CFT}
\begin{align}
 \Tr \left[ a^{t \partial} e^{b \partial} e^{c t} \right]
 & =
 \frac{1}{1-a} \exp \left( \frac{abc}{1-a} \right)
 \label{eq:abc_formula}
\end{align}
because
\begin{align}
 \frac{1}{\pi} \int d^2 z \, e^{-|z|^2}
 (z| a^{t\partial_t} e^{bt} e^{c\partial_t} |z)
 & =
 \frac{1}{\pi} \int d^2 z \,
 e^{-(1-a)|z|^2 + ab z^* + cz}
 \, .
\end{align}

\subsection{Torus correlation function}

Let us compute the torus correlation function~\eqref{eq:SS-pair-6d}.
The product of the 5d screening currents is given by
\begin{align}
 S_{i,x}^\text{5d} S_{j,x'}^\text{5d}
 & =
 \exp
 \left(
 - \sum_{m=1}^\infty \frac{1}{m} \frac{1-q_1^m}{1-q_2^{-m}}
 c_{ji}^{[m]} 
 \frac{x'^m}{x^m}
 \right)
 : S_{i,x}^\text{5d} S_{j,x'}^\text{5d} :
 \, .
\end{align}
Then we  compute the trace part
\begin{align}
 \Tr
 \left[
  p^{L_0} : S_{i,x}^\text{5d} S_{j,x'}^\text{5d} :
 \right]
 & =
 \Tr \Bigg[
 \left(
 \prod_{i' \in \Gamma_0}
 \prod_{n = 1}^\infty p^{n t_{i',n} \partial_{i',n}}
 \right)
 \exp \left(
  \sum_{n=1}^\infty (1-q_1^n) \left(x^n t_{i,n} + x'^n t_{j,n}\right)
 \right)
 \nonumber \\
 & \qquad \times
 \exp \left(
 \sum_{n=1}^\infty - \frac{1}{n(1-q_2^{-n})}
 \left(
 x^{-n} c_{ki}^{[n]} 
 \partial_{k,n}
 + x'^{-n} c_{lj}^{[n]} 
 \partial_{l,n}
 \right)
 \right) \Bigg]
 \nonumber \\ 
 & =
 \exp
 \left(
 \sum_{n=1}^\infty
 \left(
 - \frac{1-q_1^n}{n(1-q_2^{-n})}
 \frac{p^n}{1-p^n}
 c_{ji}^{[n]} 
 \frac{x'^n}{x^n} 
 +
 \frac{1-q_1^{-n}}{n(1-q_2^{n})}
 \frac{1}{1-p^{-n}}
 c_{ji}^{[-n]} 
 \frac{x^{n}}{x'^{n}}
 \right)
 \right)
 \nonumber \\
 & \qquad \times \text{const} 
\end{align}
where we have used the formulas \eqref{eq:abc_formula} and \eqref{eq:Cartan_inv}, and the constant term does not contain $x$ nor $x'$.
Thus we obtain the torus correlator
\begin{align}
 \Tr
 \left[
  p^{L_0} S_{i,x}^\text{5d} S_{j,x'}^\text{5d}
 \right]
 & =
 \exp
 \left(
 - \sum_{n \neq 0} \frac{1-q_1^n}{n(1-q_2^{-n})(1-p^n)}
 c_{ji}^{[n]} 
 \frac{x'^n}{x^n} 
 \right)
 \, .
\end{align}
This is equivalent to \eqref{eq:SS-pair-5d}, and proves the relation \eqref{eq:SS-paier-eq}.


\bibliographystyle{../wquiver/utphysurl}
\bibliography{../wquiver/wquiver}

\end{document}